\documentstyle[12pt]{article}
\begin{document}

\author{}
\title{{\Large {\bf Excited Binomial States and Excited 
Negative Binomial States of
the Radiation Field and Some of their Statistical Properties }}}
\maketitle

\begin{center}
{\bf Xiao-Guang Wang}$^a$\footnote{%
e-mail:xyw@aphy.iphy.ac.cn} and {\bf Hong-Chen Fu}$^b$\\[0pt]

$a$ Laboratory of Optical Physics,Institute of Physics,Chinese Academy of
Sciences,Beijing,100080,P.R.China\\[0pt]
$b$ Institute of Theoretical Physics, Northeast Normal
University,Changchun,130024,P.R.China\\
\end{center}
\vspace{0.2cm}
\baselineskip=20pt
{\bf Abstract:}We introduce excited binomial
states and excited negative binomial states of
the radiation field by repeated
application of the photon creation operator on binomial states
and negative binomial states. They reduce to Fock states and
excited coherent
states in certain limits and can be viewed as intermediate
states between Fock states and coherent states.
We find that both the excited binomial states and
excited negative binomial states can be exactly normalized in terms of 
hypergeometric functions. Base on this interesting character, some of the
statistical properties are discussed. \newline

\newpage
\section{Introduction}   
\baselineskip=18pt
The interesting non-classical states which are
engendered by excitations on particular quantum states have been examined.
For the first time these states were introduced by Agarwal and Tara
as the excited coherent states (ECS)[1].
The ECS exhibit remarkable non-classical properties
such as sub-Poisonnian photon statistics and
squeezing in one of the quadratures of the radiation field,
etc. Several other excited quantum states, even and odd ECS[2], excited
squeezed states [3-6] and excited thermal states[7,8] have been discussed in the
literature. It was shown that some of the excited quantum states can be prepared in
the processes of the field-atom interaction in a cavity[1] or via
conditional measurement on a beam splitter[9].

In this work, we introduce the excited binomial states (EBS) and excited
negative binomial states (ENBS) of the radiation field
by repeated application of the
photon creation operator on binomial states (BS)[10-15] and negative
binomial states (NBS)[16-22]. We examine the mathematical and physical
properties of such states. They interpolate between Fock states and
coherent states in the sense that they reduce to Fock states and coherent
states in different limits. An interesting property is found that both the
EBS and ENBS are exactly normalized in terms of hypergeometric functions[23].
These are demonstrated in section 2. In section 3, we study the
sub-Poissonian statistics and squeezing effects of these states.
The conclusion is given in section 4.

\section{EBS and ENBS}

We introduce the EBS $|k,\eta ,M\rangle $ and
ENBS $|k,\eta ,M\rangle ^{-}$ defined by

\begin{eqnarray}
|k,\eta ,M\rangle &=& {\cal N} (k,\eta ,M)a^{\dagger k}|\eta ,M\rangle , \\
|k,\eta ,M\rangle ^{-} &=&{\cal N}^{-}(k,\eta ,M)a^{\dagger k}|\eta
,M\rangle ^{-},  
\end{eqnarray}
where 
\begin{eqnarray}
|\eta ,M\rangle &=&\sum_{n=0}^M C_n(\eta,M)|n\rangle
=\sum_{n=0}^M {M \choose n}^{1/2}\eta ^n(1-\eta ^2)^{(M-n)/2}|n\rangle ,   \\
|\eta ,M\rangle ^{-} &=&\sum_{n=0}^\infty C_n^-(\eta,M)|n\rangle
=\sum_{n=0}^\infty {M+n-1 \choose n}^{1/2}\eta ^n(1-\eta ^2)^{M/2}|n\rangle
\end{eqnarray}
are the BS and NBS, respectively. Here $k,n$ and $M$ are integers,
$\eta $ is real number, $|n\rangle $ is Fock state ,
$a^{\dagger }$ and its conjugate $a$
are boson creation and annihilation operators, ${\cal N}(k,\eta ,M)$ and $%
{\cal N}^{-}(k,\eta ,M)$ are normalization constants of the EBS and ENBS
 ,respectively.

In two limits, $\eta \rightarrow 1$ and $\eta \rightarrow 0,$
the BS reduce to
Fock state $|M\rangle =|1,M\rangle $ and vacuum state $|0\rangle
=|0,M\rangle $, and the EBS to Fock states $|M+k\rangle $ and $|k\rangle $,
respectively. In a different limit of $M\rightarrow \infty ,\eta \rightarrow
0$ with $\eta ^2M=\alpha ^2$ fixed($\alpha$ is real) , the BS reduce to coherent
states $|\alpha \rangle $ and
the EBS to the ECS. 
In the limit of $\eta \rightarrow 0,$ the NBS reduce to vacuum
state and ENBS to the Fock state $|k\rangle .$ When $M\rightarrow \infty
,\eta \rightarrow 0$ with $\eta ^2M=\alpha ^2$ fixed, the NBS reduce  to coherent
states and the ENBS to the ECS.
Thus, both the EBS and ENBS can be reduced to the Fock states
and ECS. Note that the ECS can be reduced to
coherent states, the two excited states
can be recognized as intermediate states between Fock states
and coherent states.

In determining the normalization constant of the EBS, we calculate the
expectation value of operator $a^ka^{\dagger k}$ on the BS, 
\begin{eqnarray}
B(k,\eta ,M) &=&\langle \eta ,M|a^ka^{\dagger k}|\eta ,M\rangle \nonumber \\ 
&=&M!(1-\eta ^2)^M\sum_{n=0}^M\frac{(n+k)!}{n!n!(M-n)!}\left( \frac{\eta ^2}
{1-\eta ^2}\right) ^n \nonumber \\  
&=&M!\eta ^{2M}\sum_{n=0}^M\frac{(M+k-n)!}{n!(M-n)!(M-n)!}\left( \frac{%
1-\eta ^2}{\eta ^2}\right) ^n  \nonumber \\    
&=&\eta ^{2M}[(M+k)!/M!]_2F_1(-M,-M;-M-k;(\eta ^2-1)/\eta ^2),  
\end{eqnarray}
where $_2F_1(\alpha ,\beta ;\gamma ;x)$ is the hypergeometric function[23].
The normalization constant ${\cal N}(k,\eta ,M)=1/\sqrt{B(k,\eta ,M)}.$

Using the same procedure as above, we give the expectation value of operator 
$a^ka^{\dagger k}$ on NBS 
\begin{eqnarray}
B^{-}(k,\eta ,M) &=&^{-}\langle \eta ,M|a^ka^{\dagger k}|\eta ,M\rangle ^{-}
\nonumber \\   
&=&\frac{(1-\eta ^2)^M}{(M-1)!}\sum_{n=0}^\infty \frac{(M+n-1)!(n+k)!}{n!n!}%
\eta ^{2n}.  
\end{eqnarray}
It is a infinite sum. The normalization constant of the ENBS ${\cal N}^{-}(k,\eta ,M)=$\\
$1/\sqrt{B^{-}(k,\eta ,M)}.$

Now we try to give a different form of $%
B^{-}(k,\eta ,M)$ as a finite sum. The normal ordering of the operator $%
a^ka^{\dagger k}$ is 
\begin{equation}
a^ka^{\dagger k}=\sum_{l=0}^k\frac{k!k!a^{\dagger (k-l)}a^{k-l}}{%
l!(k-l)!(k-l)!}.
\end{equation}
It is easy to evaluate from Eq.(4) that 
\begin{eqnarray}
a^k|\eta ,M\rangle ^{-} &=&\left( \frac \eta {\sqrt{1-\eta ^2}}\right) ^k%
\sqrt{\frac{(M+k-1)!}{(M-1)!}}|\eta ,M+k\rangle ^{-} \\
^{-}\langle \eta ,M|a^{\dagger k}a^k|\eta ,M\rangle ^{-} &=&\left( \frac{%
\eta ^2}{1-\eta ^2}\right) ^k\frac{(M+k-1)!}{(M-1)!}.
\end{eqnarray}

From Eq.(7) and Eq.(9), we have
\begin{equation}
B^{-}(k,\eta ,M)=\sum_{l=0}^k\frac{k!k!(M+k-l-1)!}{l!(k-l)!(k-l)!(M-1)!}%
\left(\frac{\eta ^2}{1-\eta ^2}\right)^{k-l}.
\end{equation}
The above equation can be written in terms of hypergeometric functions as 
\begin{eqnarray}
B^{-}(k,\eta,M)&=&\left( \frac{\eta ^2}{1-\eta ^2} \right)^k\frac{(M+k-1)!}{(M-1)!}\\ \nonumber  
&& _2F_1(-k,-k;-M-k+1;(\eta ^2-1)/\eta ^2).
\end{eqnarray}

It is interesting that the normalization constants of both
the EBS and ENBS can be
expressed in terms of hypergeometric function. The expressions are useful in
the following investigations of non-classcal properties
of the two excited states.

%%%%%%%%%%%%%%%%%%%%%%%%%%%%%%%%%%%%%%%%%%%%%%%%%%%%%%%%%%%%%%%%%%%
\section{Non-classical properties}

We expand the EBS and ENBS in terms of Fock states as 
\begin{eqnarray}
|k,\eta ,M\rangle &=&\sum_{n=k}^{M+k}D_n(k,\eta ,M)|n\rangle \nonumber \\
&=&{\cal N}(k,\eta ,M)\sum_{n=k}^{M+k}C_{n-k}(\eta ,M)\sqrt{\frac{n!}{%
(n-k)!}}|n\rangle   \\
|k,\eta ,M\rangle ^{-} &=&\sum_{n=k}^\infty D_n^{-}(k,\eta ,M)|n\rangle
\nonumber \\
&=&{\cal N}^{-}(k,\eta ,M)\sum_{n=k}^\infty C_{n-k}^{-}(\eta ,M)\sqrt{%
\frac{n!}{(n-k)!}}|n\rangle .  
\end{eqnarray}
It can be seen that the Fock states $|0\rangle ,|1\rangle ,...,|k-1\rangle $
are removed from the BS and NBS.

\subsection{Photon statistics}%%%%%%%%%%%%%%%%%%%%%%%%%%

The mean photon number of the EBS are given by 
\begin{eqnarray}
\langle k,\eta ,M|a^{\dagger }a|k,\eta ,M\rangle
&=&\langle k, \eta ,M|aa^{\dagger
}|k,\eta ,M\rangle -1   \nonumber \\
&=&\frac{B(k+1,\eta ,M)}{B(k,\eta ,M)}-1.  
\end{eqnarray}
The mean value of the operator $(a^{\dagger }a)^2$ can be
obtained by expressing it in the antinormally ordered form $%
(a^{\dagger }a)^2=a^2a^{\dagger 2}-3aa^{\dagger }+1$ as 
\begin{equation}
\langle k,\eta ,M|(a^{\dagger }a)^2|k,\eta ,M\rangle =\frac{B(k+2,\eta ,M)}{%
B(k,\eta ,M)}-3\frac{B(k+1,\eta ,M)}{B(k,\eta ,M)}+1.
\end{equation}
Mandel's Q parameter defined by 
\begin{equation}
Q=\frac{\langle (a^{\dagger }a)^2\rangle -\langle a^{\dagger }a\rangle ^2}{%
\langle a^{\dagger }a\rangle }-1
\end{equation}
measures the deviation from the Poisson distribution (the coherent states, $%
Q=0$). If $Q<0(>0)$, the field is called sub(super)-Poissonian,
respectively.

From Eq.(14),(15) and (16), we can easily derive Mandel's Q parameter
\begin{eqnarray}
Q &=&[B(k+2,\eta ,M)B(k,\eta ,M)-B(k+1,\eta ,M)B(k,\eta ,M)-  \nonumber \\
&&B(k+1,\eta ,M)^2]/[B(k+1,\eta ,M)B(k,\eta ,M)-B(k,\eta ,M)^2]-1.
\end{eqnarray}
Let $B\rightarrow B^{-}$ in Eq.(14), (15) and (17), we can obtain
the expectation values of $a^{\dagger
}a,(a^{\dagger }a)^2$ on the ENBS and the corresponding $Q$ parameter.

In Fig.3 we plot the $Q$ parameter of the EBS as a function
of $\eta $ for different values of $k.$ For $k=0,$ the EBS become the
BS and $
Q=-\eta^2$[13]. $Q$ parameters of the EBS $|k,0,M\rangle $ and $|k,1,M\rangle 
$ are equal to $-1$ for $k\neq 0$.
This is easily understood for the states $|k,\eta
,M\rangle $ reduce to Fock states $|k\rangle $ and $|k+M\rangle $ in the
limits $\eta \rightarrow 0$ and $\eta \rightarrow 1$ and the $Q$ parameter
of Fock states $|n\rangle (n\neq 0)$ are equal to $-1$.
We see that the field
in the EBS exhibit a significant amount of
sub-Poissonian statistics.
The $Q$ parameters of the ENBS are displayed in Fig.2
as a function of $\eta $ for different values of $k.
 $For $k=0,$ the ENBS becomes
the NBS. The $Q$ parameter of the NBS is equal to $\eta^2 /(1-\eta^2 )[20]$
and the NBS
always show super-Poissonian statistics except for $\eta =0.$ For $k\neq 0$
and $\eta\rightarrow 0$, the ENBS $|k,0,M\rangle ^{-}$ becomes Fock states $|k\rangle $ and $Q=-1.$ Like
the EBS, the ENBS also shows a significant amount of sub-Possonian statistics. From
Fig.1 and Fig.2, we  find that the sub-Poissonian statistics of both the EBS
and ENBS are enhanced as $k$ increases.

\subsection{Squeezing effects}%%%%%%%%%%%%%%%%%%%%%%%%%%5

From Eq.(12) and (13), the expectation values of operator $a$ and $a^2$ on
the EBS and ENBS are obtained as
\begin{eqnarray}
\langle k,\eta ,M|a|k,\eta ,M\rangle  &=&\sum_{n=k}^{M+k-1}\sqrt{n+1}D_n(k,\eta
,M)D_{n+1}(k,\eta ,M) \nonumber \\ 
\langle k,\eta ,M|a^2|k,\eta ,M\rangle  &=&\sum_{n=k}^{M+k-2}\sqrt{(n+2)(n+1)}%
D_n(k,\eta ,M)D_{n+2}(k,\eta ,M)  \nonumber \\
^{-}\langle k,\eta ,M|a|k,\eta ,M\rangle ^{-} &=&\sum_{n=k}^\infty \sqrt{n+1}%
D_n^{-}(k,\eta ,M)D_{n+1}^{-}(k,\eta ,M)  \nonumber \\
^{-}\langle k,\eta ,M|a^2|k,\eta ,M\rangle ^{-} &=&\sum_{n=k}^\infty \sqrt{%
(n+2)(n+1)}D_n^{-}(k,\eta ,M)D_{n+2}^{-}(k,\eta ,M).  
\end{eqnarray}
Define the quadrature operators $x$ (coordinate) and $p($momentum$)$ by 
\begin{equation}
x=\frac 1{{2}}(a^{\dagger }+a),p=\frac i{{2}}(a^{\dagger }-a)
\end{equation}
In the present case $\langle a\rangle $ and $\langle a^2\rangle $ are
real. Thus, the variances of $x$ and $p$ are 
\begin{eqnarray}
Var(x) &=&\frac 14+\frac 12(\langle a^{\dagger }a\rangle +\langle a^2\rangle
-2\langle a\rangle ^2), \\
Var(p) &=&\frac 14+\frac 12(\langle a^{\dagger }a\rangle -\langle a^2\rangle
).  \nonumber
\end{eqnarray}
Combining Eq.(14),(18) and (20), we can study the squeezing effects of the EBS and
ENBS.

In Fig.3 we show the quantity $Var(x)$ of the EBS as a function of $\eta $ for
different values of $k$ .The squeezing occurs in the
quadrature component $x$ over a wide range of
parameters. As $k$ increases, the range of squeezing becomes narrow and the
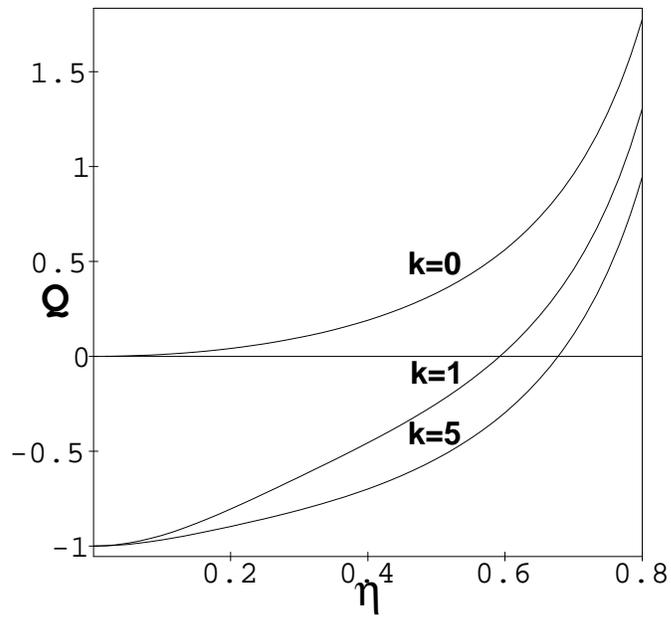
degree of squeezing is reduced. Fig.4 gives the quantity $Var(p)$ of the ENBS
as a function of $\eta $ for different values of $k.$ The ENBS  exhibit
squeezing in the quadrature component $p$ .As the EBS, the range of squeezing becomes narrow and the degree of
squeezing is reduced as $k$ increases.

\section{Conclusions}

We have introduced and investigated the EBS and ENBS
and found the following:

(1) The EBS and ENBS can be reduced to the Fock states and 
ECS. As intermediate states, both the EBS and ENBS
interpolate between the Fock states and coherent states.

(2)The EBS and ENBS are exactly normalized in terms of hypergeometric
functions.

(3)The  EBS always shows sub-Poissonian statistics.
NBS always shows super-Poissonian statistics, but the ENBS
can be sub-Poissonian. The sub-Poissonian statistics are enhanced
as $k$ increases for both the EBS and ENBS. In contrary
to this, the squeezing properties of the two excited states
are reduced and the range
of squeezing becomes narrower as $k$ increases. The squeezing occurs in the
quadrature component $x$ of the EBS and
in the quadrature component $p$ of the ENBS
over a wide range of parameters.

The BS and NBS can be generated in some nonlinear processes[11,12,20]. Then, by a
similar production mechanism of the ECS[1], let the excited atoms pass through a
cavity and provide the field in the cavity is initially in a binomial state
or a negative binomial state, one can produce the EBS and ENBS.

{\bf Acknowledgement:} This work is supported in part by the National Science
Foundation of China.

\end{document}